# Research Method Usage across Academic Ages in Library and Information Science: An Empirical Study (1990–2023)

Chengzhi Zhang, Jiayi Hao, Yi Mao

Department of Information Management, Nanjing University of Science and Technology, Nanjing, 210094, China

Corresponding author: Chengzhi Zhang (zhangcz@njust.edu.cn).

**Abstract:** Academic age critically shapes career development, influencing research behavior, output volume, and methodological choices. Analyzing method variation across academic ages offers a new theoretical lens on scholarly evolution and provides early-career researchers with practical guidance for method selection. A corpus of 26,677 articles published 1990–2023 in 14 authoritative Library and Information Science journals was compiled. The CogFT model automatically classified the research methods embedded in these articles, and Top2Vec generated the topic model. This process resulted in a comprehensive dataset linking research methods with topics. Author-name disambiguation enabled calculation of each scholar’s academic age. Popularity and Shannon diversity indices for methods, together with topic diversity, were compared across academic age groups. Results reveal dynamic methodological trends: the share of theoretical approaches declined gradually, whereas experimental and bibliometric methods gained ground. Method popularity differs significantly among cohorts. Mid-career scholars exhibit the highest method diversity; late-career scholars the lowest.

**Keywords:** Research Methods; Academic Age; Research Topics; Library and Information Science (LIS)

## 1. Introduction

In an era where knowledge innovation drives sustainable social development, researchers are central to knowledge production and dissemination. Their scholarly characteristics—including research behavior, career trajectories, and influencing factors—have become a major focus for both academia and policymakers (Jia et al., 2017; Fortunato et al., 2018). Existing research has largely examined scholars from a macro perspective, emphasizing collective patterns (Gingras et al., 2008; Zhang et al., 2025). However, recent studies increasingly shift toward micro-level analysis, exploring individual differences throughout academic careers (Abramo et al., 2011; Kwiek & Roszka, 2022). Within this trend, "academic age" has emerged as a key lens for capturing individual scholarly development (Wang et al., 2017). It serves as a vital indicator of a researcher's career stage. Academic age helps trace the evolution of cognitive capabilities and publication patterns over time. It also offers insights into shifts in research preferences, resource acquisition, and scholarly impact.

Research methods are fundamental tools for scholarly inquiry. They reflect a researcher's expertise and the quality of their work. More importantly, methods play a crucial role in the production, dissemination, and innovation of scientific knowledge. Consequently, they serve as a significant vantage point for studying scholar characteristics. Numerous studies have investigated research methods. In Library and Information Science (LIS), Järvelin and Vakkari (1990) pioneered this line of inquiry by analyzing the use of research methods in the field through content analysis. Building upon their classification framework, Chu and Ke (2017) further developed the theoretical foundation by identifying and analyzing 16 research methods from a data collection perspective.

Existing studies have examined variations in collaboration, performance, and content preferences among scholars of different academic ages (Packalen & Bhattacharya, 2019; Simoes & Crespo, 2020; Wang et al., 2017; Zeng et al., 2019). Research has also systematically reviewed classification frameworks and usage trends of research methods (Zhang et al., 2023). Therefore, academic age, research methods, and topics are integrated here to portray scholar characteristics. Such integration refines and potentially innovates methodological practice across the LIS field.

## 2. Problem statement

Prior research has identified age-related differences in method preferences across academic career stages. Late-career scholars often prefer qualitative methods, while early-career researchers tend to employ quantitative approaches (Lou et al., 2021). Examining these differences in the LIS field is crucial. It allows for a systematic understanding of how methodological choices evolve throughout a scholarly career. Furthermore, such inquiry provides valuable guidance for early-career scholars. It helps them make informed methodological decisions at the outset of their careers, thereby enhancing their research efficiency.

Taking the LIS field as an example, this study processes data from 14 authoritative journals in the Q1-Q3 tiers of the Web of Science Journal Citation Reports (JCR). It constructs a dataset of research methods and performs topic modeling. Subsequently, author name disambiguation and academic age calculation are conducted. The aim is to investigate differences in method usage among scholars of different academic age groups from multiple dimensions, including popularity and diversity of research methods. Specifically, this study addresses the following research questions:

***RQ1***: What differences exist in the use of research methods among scholars of different academic ages in the LIS field?

***RQ2***: What characteristics are observed in the methods used by scholars across various academic age groups within different research topics?

By addressing these questions, this research aims to reveal method disparities in the scientific activities of scholars at different career stages. The findings are expected to provide a novel theoretical perspective for understanding method selection patterns throughout an academic career. Furthermore, they offer valuable insights for supporting scholar development and informing the evolving landscape of modern science.

## 3. Literature review

### *3.1 Research on academic age*

Individual characteristics such as gender and age are crucial for understanding scholars' research behavior and career trajectories. While chronological age offers insights into lifespan and energy levels, obtaining systematic data on birth years is often challenging. Consequently, academic age has been widely adopted as a proxy for investigating scholarly development (Bu et al., 2018; Győrffy et al., 2020; Sugimoto et al., 2016). Research by Kwiek and Roszka (2022) suggests that academic age serves as a better representative of chronological age in STEMM (Science, Technology, Engineering, Mathematics, and Medicine) fields compared to the social sciences and humanities. There is no single standard for calculating academic age. Two primary methods exist: using the date of a scholar's first publication or using the date of their first publication after earning a PhD. Costas et al. (2015) evaluated both approaches, finding that while both PhD graduation year and birth year correlate with the first publication date, the correlation is stronger with PhD graduation year. This evidence supports the rationale of using a scholar's first publication date to measure academic age.

### *3.2 Research on methods in LIS*

Research methods are a core element of disciplinary development, and their standardization and maturity signify the field's progress. Scholars have primarily established classification systems for research methods by analyzing journal articles or dissertations. A foundational contribution is the classification system proposed by Järvelin and Vakkari (1990), which includes 13 research methods. Subsequently, Blake (1994) analyzed LIS dissertation abstracts from 1975-1989 and proposed a multi-dimensional classification system. This system encompasses methods such as descriptive research, case studies, bibliography, historical-biographical methods, surveys (questionnaires and interviews), bibliometrics, content analysis, modeling, quasi-experimental, experimental, theoretical, and combined methods. Furthermore, Chu and Ke (2017) coded the methods used in articles published between 2001 and 2010 in three leading LIS journals. They identified and summarized 16 primary research methods, including bibliometrics, content analysis, and the Delphi study. This classification has become one of the most widely used systems in the LIS field.

Furthermore, scholars have analyzed research methods in the LIS field based on journal article data, enriching the discipline's toolkit through examinations of method evolution, diversity, and popularity. For instance, Chu (2015) analyzed the types and prevalence of methods in 1,162 articles published between 2001 and 2010, identifying content analysis, experimental methods, and theoretical research as widely used during that period. Ma & Lund, (2021) investigated the evolution and shift of research topics and methods in library and information science, revealing how methodological preferences have evolved alongside changing research themes. Subsequently, Järvelin and Vakkari (2022) provided a comprehensive longitudinal analysis of articles from over 30 LIS journals from 1965 to 2015, systematically summarizing the evolutionary trends of research methods across five decades. Vakkari et al. (2023) further extended this line of inquiry by examining how scholars' disciplinary backgrounds are associated with the evolution of topics and methods in LIS research from 1995 to 2015, highlighting the joint effects of individual attributes and thematic development on methodological choice.

Building on existing research method frameworks and manually annotated datasets, scholars have developed techniques to automatically identify methods used in academic papers. Eckle-Kohler et al. (2013) developed an automatic classification system based on multi-label classification algorithms, which could effectively identify quantitative empirical methods and other empirical research approaches. Subsequently, Zhang and Zhang (2021) proposed a semi-automatic classification approach based on clustering word vectors of method-related entities. By clustering these entities extracted from full-text articles, they expanded a traditional basic method taxonomy, constructing a more detailed hierarchical system which was successfully applied to classify methods in Chinese information science. Furthermore, Zhang et al. (2021) analyzed 2,599 articles from three LIS journals. Using a combination of content analysis and text mining, they labeled the articles with 29 distinct research methods and also experimented with a rule-based automatic identification approach. Collectively, these studies, which train classification models on established frameworks and manual annotations, explore new algorithms and technical pathways. This ongoing work advances the development of automatic method classification in LIS. These contributions not only facilitate the standardized analysis and systematic categorization of research methods but also lay a solid foundation for the present study's goal of identifying methods used by scholars of different academic ages in the LIS field.

# 4. Data and methods

## *4.1 Data collection and preprocessing*

To ensure data quality, this study selected widely recognized high-quality journals in the LIS field. The journal selection process was informed by the work of Järvelin and Vakkari (1993, 2022), who systematically identified 31 core LIS journals based on thematic analysis. This list was cross-referenced with the core journal list from the 2022 Journal Citation Reports (JCR) under the "Information Science & Library Science" category, specifically including journals ranked within Q1 to Q3. Through this process, 14 English-language academic journals were finalized as the primary data sources. To capture the longitudinal development of the field, the publication period was set from 1990 to 2023. The authoritative status and disciplinary representativeness of these selected journals provide a solid foundation for data reliability.

The collected data includes both bibliographic metadata and full-text articles. To maintain relevance and accuracy, only research articles were collected; non-research items such as book reviews, news, and editorials were excluded due to their lack of a defined research methodology. Ultimately, the dataset comprises 26,677 research articles, including full text and metadata, published between 1990 and 2023 from the 14 target journals. The distribution of articles across these journals is shown in Figure A.1 of Appendix A.

## *4.2 Research topic identification and automatic method classification*

### *4.2.1 Identification of research topics in academic papers*

This study employed the Top2Vec model (Angelov, 2020; Zhang et al., 2025) to perform topic modeling on the corpus of academic papers. After cleaning and preprocessing the abstracts, Doc2Vec maps each text to a point in a high-dimensional space; dense regions are identified as latent topics, and the centroid word vectors supply the key terms that label each topic.

To evaluate the quality of different topic cluster numbers, the $C_v$ coherence score (Röder et al., 2015) was used. As shown in Figure A.2 in Appendix A, the C_v score peaked when the number of topic clusters was set to 29, indicating the optimal initial clustering result for our dataset. The Top2Vec model output high-frequency keywords for each of these 29 initial topics and assigned each paper to a corresponding topic. During manual review and summary of the high-frequency

keywords of each initial topic, we found that some topics had obvious semantic overlap or inclusion relationships. To enhance topic generality and streamline the analysis, we merged these semantically related topics based on the consistency of their core keywords , resulting in a final set of 20 distinct research topics (see Table B.1, Table B.2 in Appendix B). These topics encompass core LIS areas such as Scientometrics and Information Retrieval, as well as emerging technical and application-oriented foci like Text Mining and Digital Information Resources Management, reflecting the breadth and diversity of the field.

### *4.2.2 Classification of research methods in academic papers*

Building upon the constructed full-text corpus, this study classified and identified the research methods used in the papers. The methodology taxonomy established by Chu and Ke (2017) served as the classification framework. The Cognize Full Text (CogFT) model (Zhang & Tian, 2023) was employed for the automated identification of research methods within the LIS academic papers.

CogFT comprises two modules: full-text encoding and method classification. First, each article is split into 128 non-overlapping segments of 64 tokens, giving a maximum input length of 8 192 tokens. SciBERT(Beltagy et al., 2019) embeds every segment; two stacked self-attention layers (eight heads each, hidden size 768) then capture inter-segment dependencies and output a probability that the segment describes the study's method. The four segments with the highest probabilities are concatenated with the first four segments of the paper to form an 8-segment"method summary."This summary is fed into a standard BERT classifier that assigns one of 16 predefined method labels. CogFT achieves an F1 score > 0.85, outperforming existing method-tagging models; it was therefore used to predict the methods employed in all 26 677 LIS articles. As a single article could be assigned multiple method labels, the total count of method instances exceeds the number of journal articles. The overall method classification for each article in the corpus is listed in Figure A.3 of Appendix A.

## *4.3 Author name disambiguation and academic age calculation*

### *4.3.1 Author name disambiguation*

To address the issue of author identity ambiguity caused by name homonyms, this study utilized the OpenAlex[1] database for name disambiguation. The unique Author ID assigned by OpenAlex was adopted as the definitive identifier to ensure accuracy and consistency. Operationally, the PyAlex library was employed to systematically process author information from the collected papers. Firstly, using each article's DOI, the OpenAlex API was called to retrieve its corresponding metadata. Secondly, the author names, affiliations, and OpenAlex Author IDs were extracted from the returned metadata for each paper. Finally, using the obtained Author IDs as core identifiers, all author records were matched and integrated to ensure distinct authors were accurately distinguished throughout the analysis. This process successfully disambiguated 53,683 authors, each assigned a unique identifier, laying the groundwork for subsequent academic age calculation and method analysis.

### *4.3.2 Author academic age calculation*

Following conventional practice, this study defines a scholar's first publication year as their academic starting point. The academic age for a given publication is calculated using Formula

[1] https://openalex.org/

(1): it is the difference between the publication year of the paper in question and the year of the author's first publication, plus one. Adding one ensures that the first publication year corresponds to an academic age of 1, thereby including the initial year in the count.

$$AAS = PYA - EPY + 1 \quad (1)$$

Where *AAS* represents the Academic Age of the Scholar, *PYA* is the Publication Year of the Article, and *EPY* is the Earliest Publication Year. Adding 1 ensures that the year of the first publication corresponds to an academic age of 1. Since the choice of research methods is typically influenced by both the first author and the corresponding author, this study analyzes the academic age data for both roles. When counting article authors in this study, for papers where the first author and the corresponding author are the same person, we count that author only once. For papers where the first author and the corresponding author are different, we count both as two distinct authors. Among the 26,677 articles, 21,876 had the same individual serving as both first author and corresponding author. To ensure data validity and reliability, outliers were identified and removed using the Interquartile Range (IQR) method.

$$Upper = Q3 + 2.2 * (Q3 - Q1) \quad (2)$$

$$Lower = Q1 - 2.2 * (Q3 - Q1) \quad (3)$$

Formulas (2) and (3) calculate the upper and lower bounds for academic age, respectively, where Q1 represents the lower quartile (25th percentile) and Q3 represents the upper quartile (75th percentile) of the academic age distribution. Due to the substantial proportion of authors with low academic ages, a coefficient of 2.2 was used instead of the common 1.5 (Chowdhary et al., 2024). This adjustment prevents the over-exclusion of authors with high academic ages as outliers. It also ensures alignment with established academic age classifications, enhancing the comparability of the findings. Values below the lower bound or above the upper bound were identified as outliers and removed. This process resulted in a final dataset of 25,669 first authors and 25,633 corresponding authors. To examine method differences across academic age groups, authors were further categorized. The 95th percentile value (47 years) was set as the maximum academic age. Based on quartiles, authors were divided into three groups: the 25th percentile (7 years) and the 75th percentile (14 years) of the distribution. Thus, authors with an academic age below 7 were defined as early-career scholars, those between 7 and 14 as mid-career scholars, and those above 14 as late-career scholars. The specific number and proportion of scholars in each group are presented in Table B.3 of Appendix B.

### *4.4 Measuring differences in research method usage among academic age groups in LIS*

This study employs two metrics—popularity and diversity—to measure differences in research method usage among scholars of different academic ages. First, the popularity of each research method is assessed using its average annual usage frequency. This measure offers a quantitative view of how frequently each method is used and helps identify shifts in method preferences within the academic community over time. The calculation for popularity is shown in Equation (4).

$$L_{m,t} = \frac{N_{m,t}}{A_t} \quad (4)$$

The popularity of a research method is operationalized as its average annual usage per author.

For a method m in year $t$, this measure, $L_{m,t}$, is defined by Equation 4, where $N_{m,t}$ signifies the total usage count of method m in year $t$, and $A_t$ represents the total number of authors in year $t$.

Second, method differences among age groups are revealed by employing Shannon entropy to measure diversity. The proportion of a method $i$ within a group, $P_i$, is calculated as $P_i = \frac{n_i}{N}$, where $n_i$ is the count of method $i$ used by the group, and $N$ is the group's total usage count of all methods.

$$H = -\sum(p_i * \ln(p_i)) \quad (5)$$

# 5. Results analysis

## *5.1 Analysis of research method usage frequency by academic age*

To address ***RQ1***, this subsection analyzes the frequency of method usage across different academic age groups. The usage frequency for each method was counted based on the methods identified for the first author and corresponding author of each article. As shown in Figure 1, the most frequently used method was bibliometrics, with a total of 7,126 instances. This was followed by the experimental method (5,461 instances), while the Delphi study was the least used (97 instances). Regarding usage patterns across scholar groups： Among early-career scholars, the experimental method was most frequent, followed by bibliometrics. Among mid-career scholars, bibliometrics was the most used method, followed by content analysis and the experimental method. Among late-career scholars, bibliometrics was also the most prevalent, followed by the experimental method. Overall, bibliometrics is the most popular and widely used method in the LIS field.

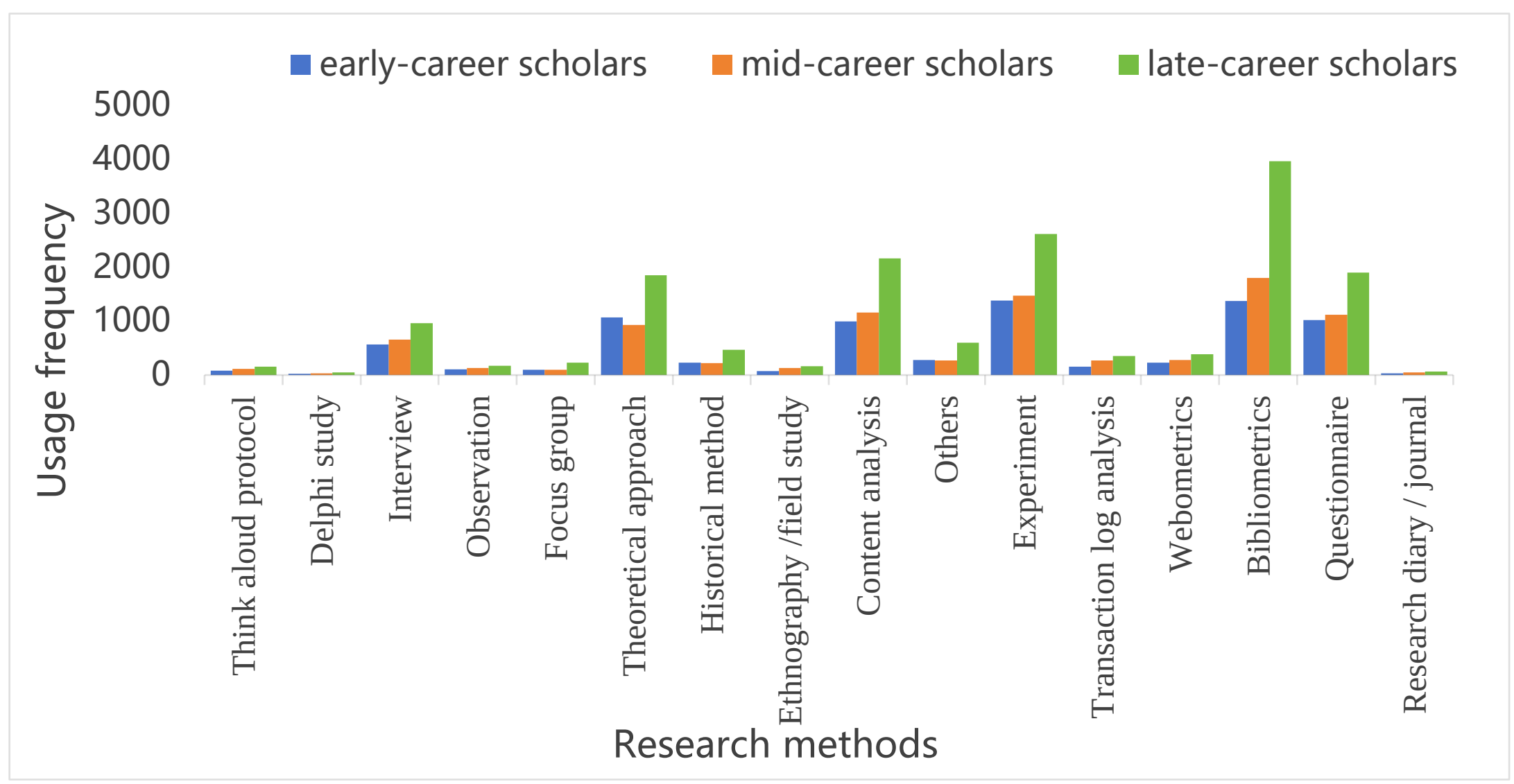

**Figure.1.** Summary of the frequency of research methods used by various types of scholars.

## *5.2 Analysis of popularity and diversity in research method usage by academic age*

### *5.2.1 Popularity of research method usage in the LIS field*

This section presents a condensed heatmap of method-share percentages by scholar group across 1990-2023. To improve readability and reduce information overload, yearly values were aggregated into three decade-based periods (1990-1999, 2000-2009, and 2010-2023). In Figure 2, darker colors indicate higher percentages. In 1990-1999, theoretical approach was dominant across all age cohorts, while experiment and bibliometrics had notable shares. In 2000-2009, method preferences became more balanced, with experiment and bibliometrics rising and theoretical approach declining. In 2010-2023, bibliometrics became the leading method across cohorts (especially among late-career scholars), with experiment ranking second, and content analysis and questionnaire remaining important.

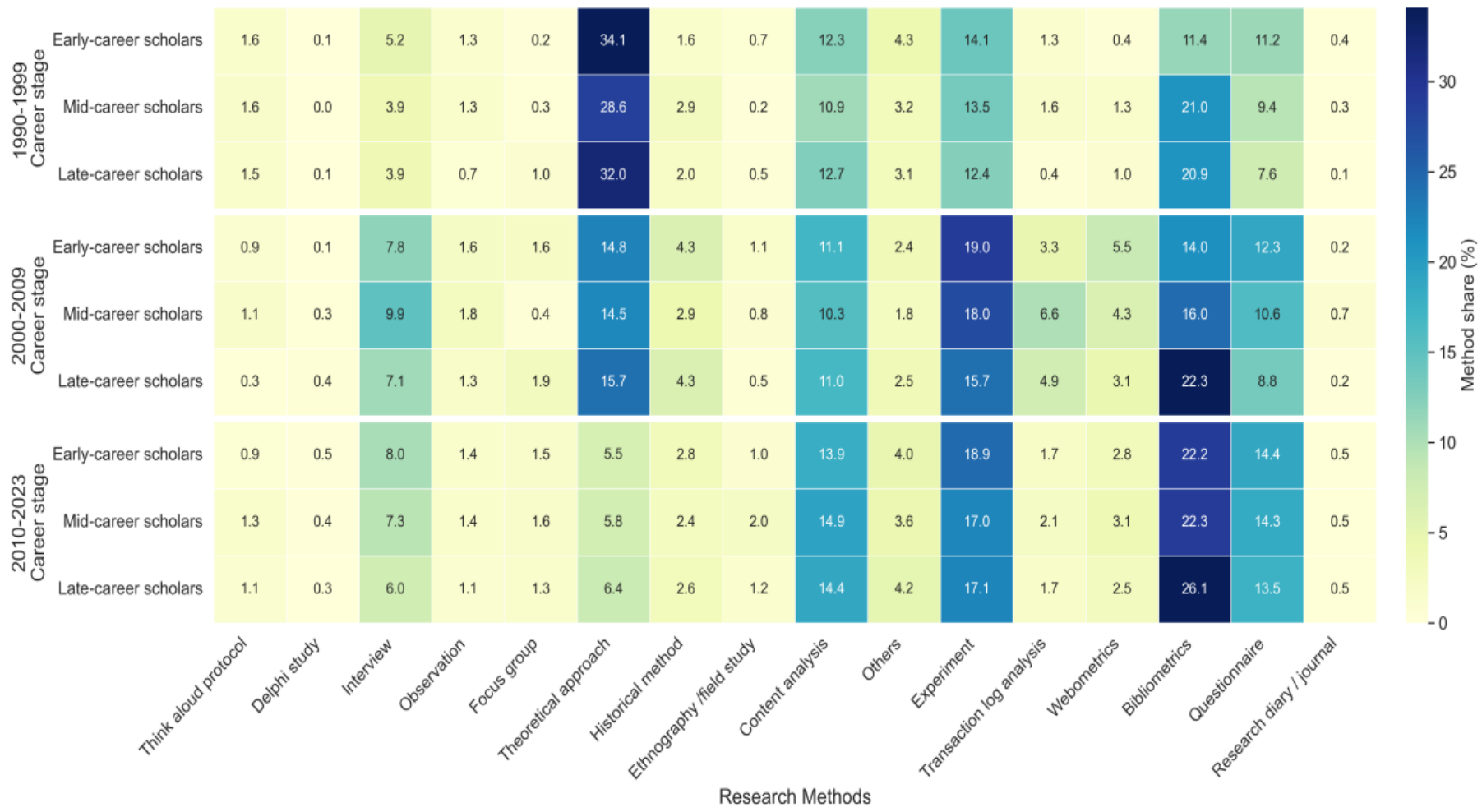

**Figure.2.** Popular research methods used by scholars across career stages.

Comparing cohorts within the three periods shows both continuity and differentiation. Across all cohorts, the relative share of theoretical approach declined markedly from 1990-1999 to 2010-2023, while bibliometrics and experiment increased. Mid-career and late-career scholars consistently exhibited higher shares of bibliometrics, whereas early-career scholars showed relatively stronger use of experiment. Content analysis and questionnaires remained stable, indicating their sustained roles as commonly used approaches.

In summary, quantitative methods (e.g., bibliometrics, experiment) and mixed methods (e.g., content analysis, questionnaire) have gradually become dominant among mid-career and late-career scholars, while the popularity of purely qualitative, theoretical approaches has steadily declined.

### *5.2.2 Diversity of research method usage in the LIS field*

**Table 1.** Diversity of research methods used.

| **Scholar Category** | **Early career Scholars** | **Mid-career Scholars** | **Late-career Scholars** |
|---|---|---|---|
| H（Shannon Entropy） | 2.269 | 2.291 | 2.218 |

Using the previously defined diversity metric, the Shannon entropy was calculated for each of the three scholar groups, as shown in Table 1. Mid-career scholars exhibited the highest entropy value, indicating the greatest diversity in their method usage, while late-career scholars displayed the lowest. To verify the statistical significance in Shannon entropy among academic age groups, we conducted a Kruskal-Wallis H test—a non-parametric method suitable for our data, with academic age as a continuous variable and research method as a categorical variable. This test avoids the assumptions of normal distribution and homogeneity of variance required by traditional ANOVA. The test yielded $H = 241.8$ ($df = 15$, $p < 0.001$), rejecting the null hypothesis and confirming significant differences in academic age distribution across the 16 research methods. This reflects a meaningful correlation between academic age and research method choice, providing empirical support for the impact of academic age on method selection.

## *5.3 Differences in research method usage by topic and academic age*

### *5.3.1Analysis of research topic diversity by academic age*

This section addresses RQ2. To investigate topic preferences across academic age groups, the distribution of publications by research topic is analyzed. As shown in Table 2, late-career scholars lead significantly in established areas like "Scientometrics" (1,620 publications) and "Text Mining" (1,061 publications), reflecting their solid foundation and substantial accumulation in these fields. In contrast, early-career scholars demonstrate strong engagement with the emerging, application-oriented topic of "Text Mining" (573 publications). Meanwhile, mid-career scholars show considerable activity in areas such as "Scientometrics" (733 publications) and "Information Search Behavior" (702 publications).

This study also assessed the diversity of research topics among authors of different academic ages using Shannon entropy. As shown in Table 3, the variation in topic diversity across academic age groups was relatively small.

**Table 2.** A summary of the frequency of research topics among various scholars.

| Research Topic | Early career Scholars | Mid-career Scholars | Late-career Scholars |
|---|---|---|---|
| Scientometrics | 456 | 733 | 1620 |
| Text Mining | 573 | 644 | 1061 |
| Information Search Behavior | 566 | 702 | 1000 |
| Knowledge Management | 499 | 546 | 973 |
| Information Retrieval | 566 | 558 | 867 |
| Information Retrieval Models and Evaluation | 459 | 547 | 931 |
| Information Theory | 350 | 543 | 1029 |
| Library Services | 562 | 501 | 813 |
| Library Education/Information Literacy | 444 | 546 | 790 |
| Bibliometric Analysis | 348 | 420 | 940 |
| International Cooperation | 406 | 408 | 879 |
| Digital Information Resources | 482 | 309 | 742 |
| Social Media/Emotion Analysis | 282 | 355 | 764 |
| Health Information Services | 305 | 395 | 696 |
| Academic Productivity | 257 | 321 | 647 |
| User Information Behavior | 251 | 334 | 543 |
| Information Dissemination | 238 | 293 | 572 |
| Data Security Management | 248 | 228 | 487 |
| Library Policy/Public Services | 311 | 197 | 427 |
| Academic Publishing | 112 | 132 | 292 |
| Total | 7715 | 8712 | 16073 |

**Table 3.** Diversity of research topics used.

| **Scholar Category** | **Early career Scholars** | **Mid-career Scholars** | **Late-career Scholars** |
|---|---|---|---|
| H（Shannon Entropy） | 2.93 | 2.92 | 2.94 |

### *5.3.2 Method preferences by research topic*

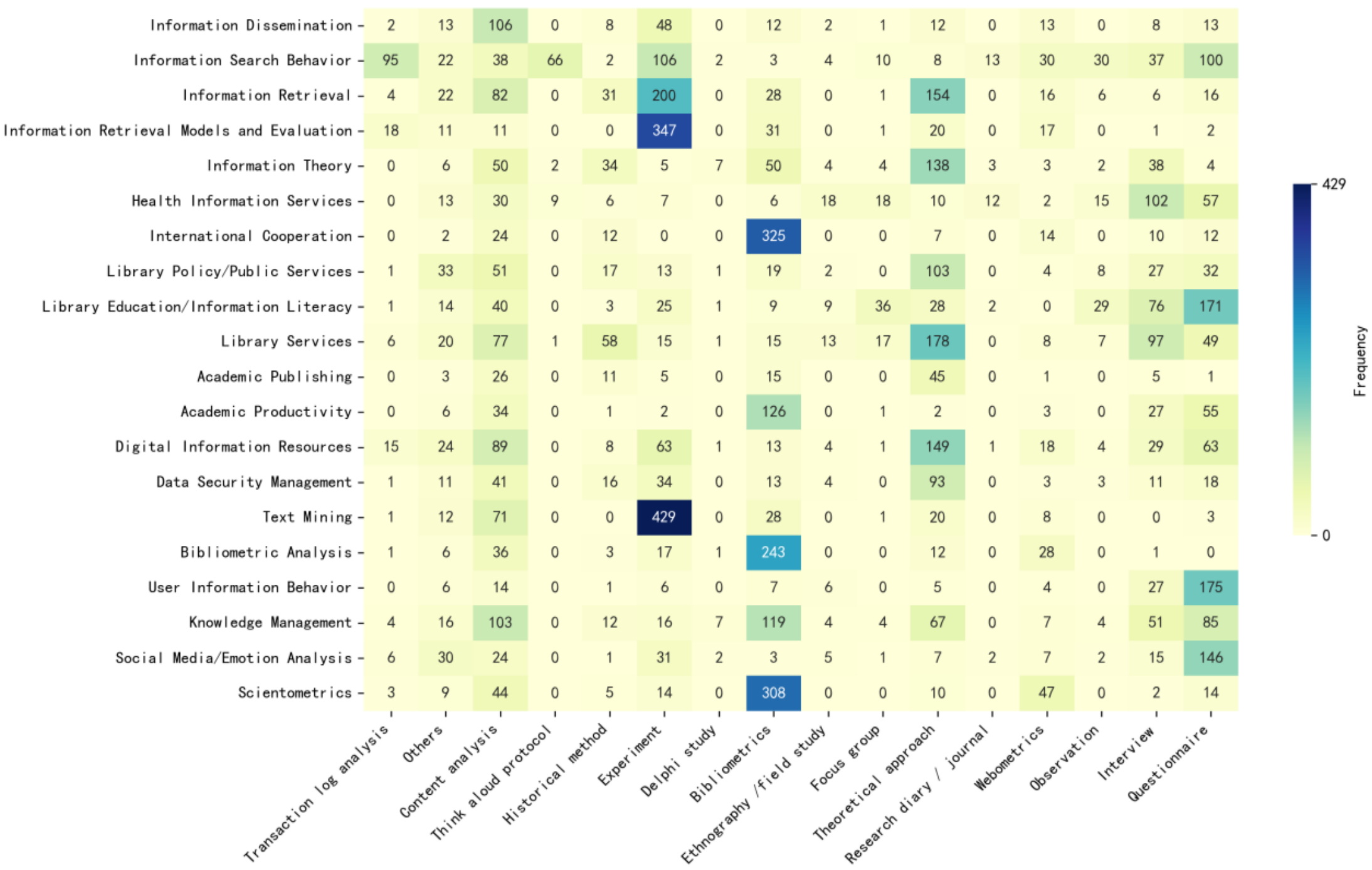

**Figure.3.** The early-career scholar's research method uses heat maps.

To investigate the characteristics and distribution of research methods across different topics, the usage frequency of 16 research methods within 20 research topics is analyzed for the three scholar groups. Figure 3 illustrates the distribution for early-career scholars. Across all years, the experimental method consistently ranks among the most frequently used approaches in every major topic area. Its usage is particularly prominent in "Text Mining" (429 instances) and "Information Retrieval Models and Evaluation" (327 instances). The bibliometric method was also frequently used across multiple topics, while the theoretical approach was widely applied in areas such as "Library Services," "Information Retrieval," and "Digital Information Resources." In contrast, the usage frequency of content analysis was relatively balanced across various topics. Laterally, the questionnaire method was the most used technique in research topics like "User Information Behavior," "Social Media/Emotion Analysis," and "Library Education/Information Literacy. "Early-career scholars prefer content analysis in the "Information Communication" topic. For "Scientometrics," they predominantly employed the bibliometric method. Figure A.4 in Appendix A shows the method distribution across research topics for mid-career scholars. In the field of Scientometrics, bibliometrics were used most frequently (557 instances). The overall color distribution in the heatmap reveals that mid-career scholars show a clear preference for three main methods: bibliometrics, experimental methods, and questionnaires. Figure A.5 in Appendix A displays the method distribution across research topics for late-career scholars. In the field of Scientometrics, late-career scholars demonstrated the most prominent reliance on bibliometrics, with 1,304 instances of usage. Additionally, late-career scholars frequently employed the questionnaire method, particularly in topics such as Social Media/Emotion Analysis, User Information Behavior, and Library Education/Information Literacy.

# 6. Discussion

## *6.1 Research implications.*

### *6.1.1 Theoretical implications*

This study enriches the research perspective on method practices in LIS. While existing method research primarily focuses on the suitability of methods for specific topics, classification frameworks, or macro-level evolutionary trends, it seldom explores the connection with individual scholar characteristics. By introducing the perspective of academic age, this study expands the theoretical framework for method inquiry and deepens the understanding of diversity in method selection. It offers a new lens for examining research methods. Furthermore, this study aligns its findings with prior longitudinal research on methodological evolution in LIS. Järvelin & Vakkari (2022) conducted a 50-year analysis of over 30 LIS journals and documented a long-term trend of increasing methodological scattering. Empirical strategies (e.g., surveys, scientometrics, experiments, case studies, qualitative methods) have gained popularity, while conceptual approaches have declined. This scattering is closely tied to changes in the discipline's topical landscape. Our findings on method preferences across academic age groups complement their work. While Järvelin & Vakkari (2022) identified macro-level methodological scattering and an empirical turn, we further show that this trend interacts with scholars' academic age. Senior scholars (academic age >14 years) tend to rely on traditional quantitative methods such as bibliometrics, whereas junior scholars are more inclined to adopt emerging empirical methods like text mining. This pattern reflects intergenerational transmission and innovation amid the ongoing methodological diversification of LIS. Finally, it enriches the content for constructing scholarly profiles. Compared to previous studies that examined factors such as gender, team composition, and academic roles, this investigation focuses on a scholar's academic age to explore method preferences. This approach adds a valuable dimension to scholarly profiling research. Finally, it extends the research dimension concerning the relationship between method choice and career development. The differences in method usage across academic age groups reflect not only preferences at different career stages but are also potentially shaped by the academic environment, disciplinary paradigms, and career progression. This perspective helps reveal the characteristics of method choice at various stages of a scholarly career, providing new theoretical support for academic career development studies.

### *6.1.2 Practical implications*

This research offers practical value in several aspects. First, it assists researchers in making informed method choices. It is important to emphasize that method choices are primarily driven by research questions, which evolve over time. Our findings are not meant to suggest that researchers should simply follow the preferences of scholars at different career stages. Rather, they serve as a reference when multiple methods are viable for a given research question. For example, both content analysis and text mining can be used to analyze academic literature. By understanding how method preferences vary across career stages, researchers can gain insights into the application contexts and strengths of different methods, helping them select the most appropriate approach for their specific research question and practical needs. Furthermore, the analysis provides suggestions for promoting method diversity within academia. Documented differences in method preferences across academic age groups can inform institutional policies aimed at encouraging a wider range of research practices and fostering method pluralism. Lastly, the findings can guide scholars in selecting suitable collaborators. By recognizing the distinct method and topical profiles associated with different career stages, researchers can more effectively identify potential partners with complementary expertise and aligned interests. This facilitates the formation of productive teams, ultimately enhancing project quality and success.

### *6.2 Limitations*

This study explores the relationship between academic age and research method usage among LIS scholars. However, due to the complexity of this relationship and the interplay of multiple influencing factors, several limitations remain. First, the data coverage is limited, as the study draws solely from 14 leading LIS journals, excluding other relevant publications in the field. Second, the exploration of causality is insufficient; while significant correlations are identified, the underlying causal mechanisms are not examined. Third, the dynamic evolution of scholars' method choices is not fully captured, as the analysis relies on static data, limiting insights into career-long developmental trajectories.

Fourth, the representativeness of our results may be affected by the interdisciplinary nature of LIS journals. As Vakkari et al. (2023) showed, most contributions to LIS journals in 2015 came from scholars outside the discipline. Since this study did not identify authors' disciplinary affiliations, some of the observed "method use patterns" may reflect the method preferences of non-LIS scholars rather than those of pure LIS researchers.

## 7. Conclusion

This study offers a new theoretical perspective for understanding how method choices evolve throughout a scholarly career. It also provides practical insights for supporting academic career development and optimizing research method education. Ultimately, it contributes empirical evidence to encourage method innovation and advancement in the LIS field. Future work will extend the dataset to include Chinese-language literature and underrepresented journals to enable cross-cultural and cross-linguistic comparisons, enhancing the generalizability of the findings. Additionally, employing causal inference methods, such as structural equation modeling (SEM) or causal effect analysis, could clarify how academic age influences method selection and whether method preferences, in turn, shape career paths. Finally, integrating longitudinal data with dynamic trajectory analysis would allow tracking individual scholars' method evolution over time, revealing critical transitions in career stages and their impact on research practices. Such an approach would provide a deeper, dynamic understanding of the complex relationship between academic age and method choice. Future research could further explore whether method preferences differ between LIS scholars and interdisciplinary scholars publishing in LIS journals.

## Acknowledgements

This work was supported by the National Natural Science Foundation of China (Grant No.72074113), Major Projects of National Social Science Fund (Grant No. 25&ZD298)

# Appendix

## Appendix. A

The distribution of articles across these journals is shown in Figure A. 1. As the results indicate, Scientometrics contributed the most articles (5,926, representing 22.2% of the dataset), while Library Quarterly contributed the fewest (502 articles, constituting less than 2%).

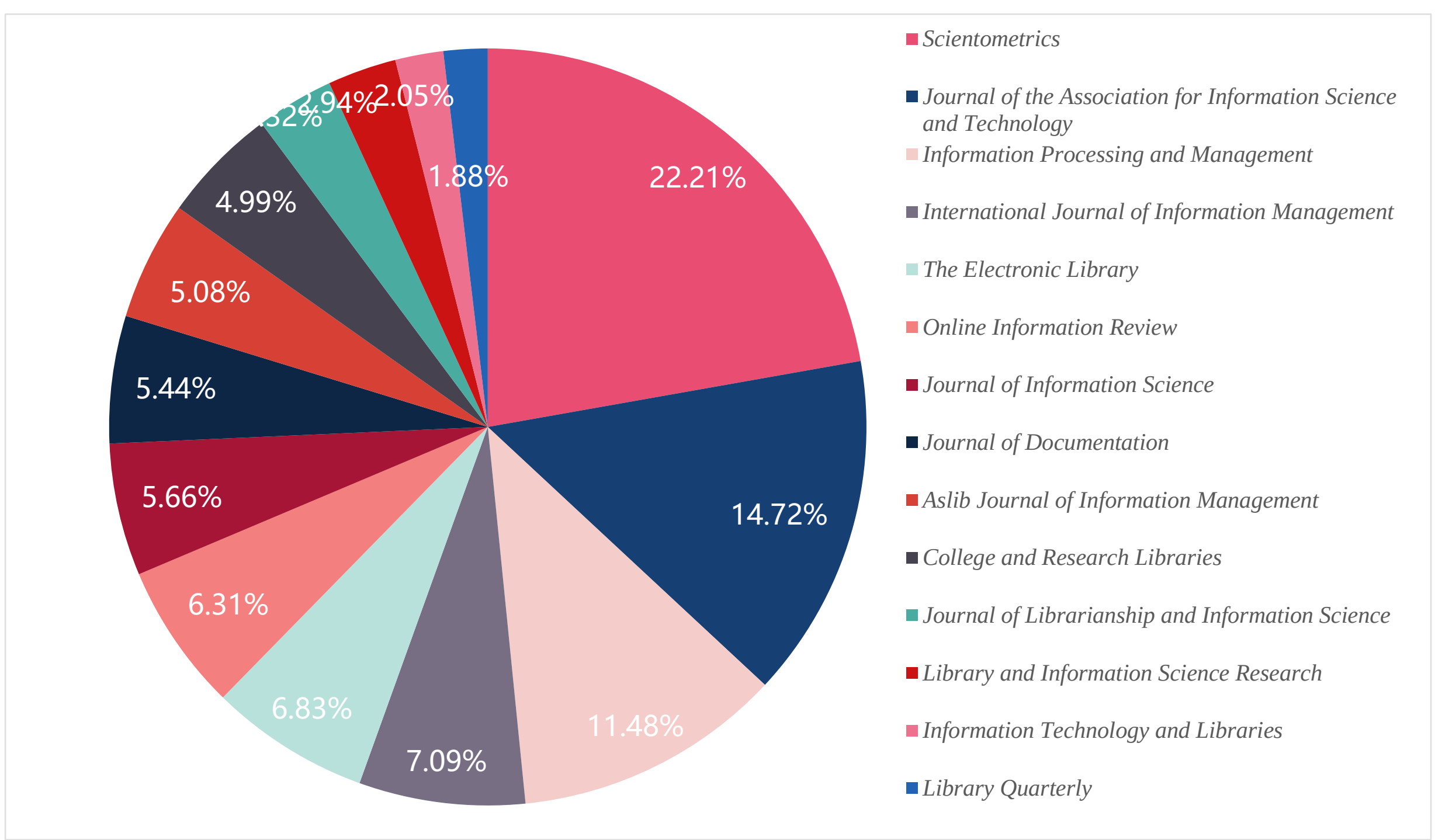

Figure A. 1. Proportion of journal paper data.

Figure A. 2 presents the overall classification results of research methods across the corpus. As a single article could be assigned multiple method labels, the total count of method instances exceeds the number of journal articles. The final tally shows that the bibliometric method was used most frequently (7,126 instances). This high frequency is likely attributable to the substantial proportion of bibliometrics-related papers within the selected journals constituting the corpus.

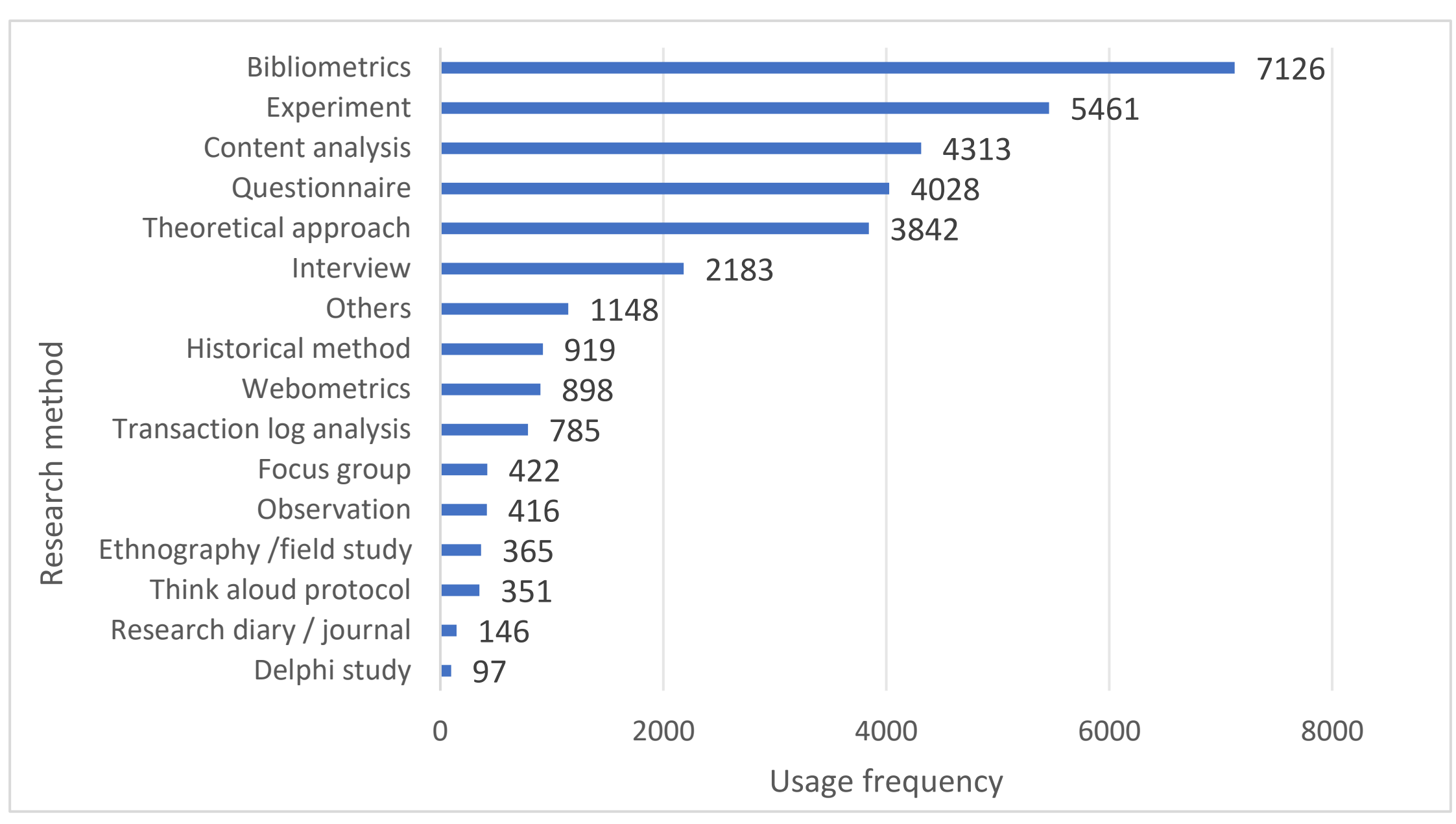

Figure A. 2. Overall frequency of research methods used.

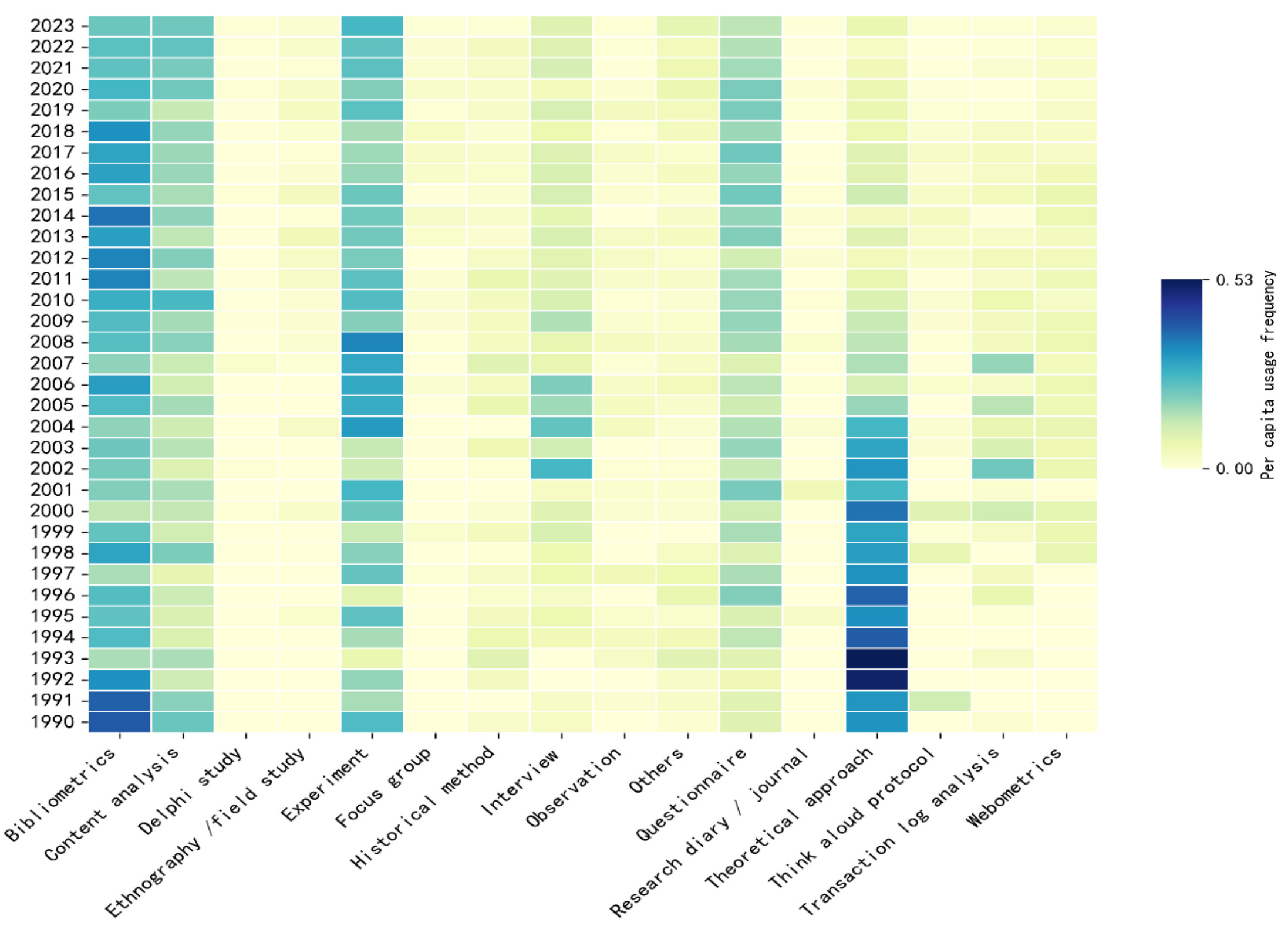

Figure A. 3. Proportion of research methods over time chart.Mid-career scholars use popular methods in their research.

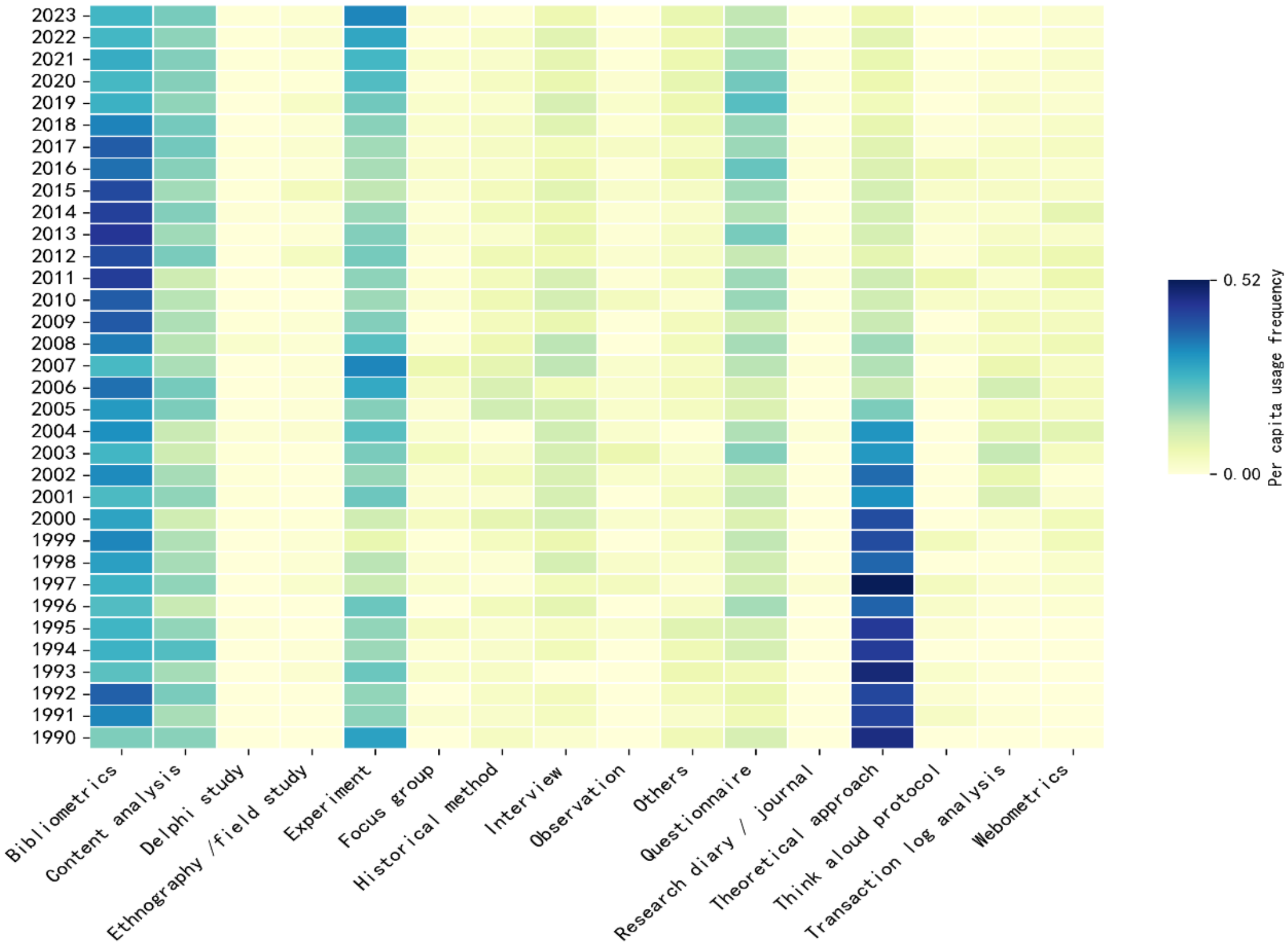

Figure A. 4. The popularity of research methods used by late-career scholars.

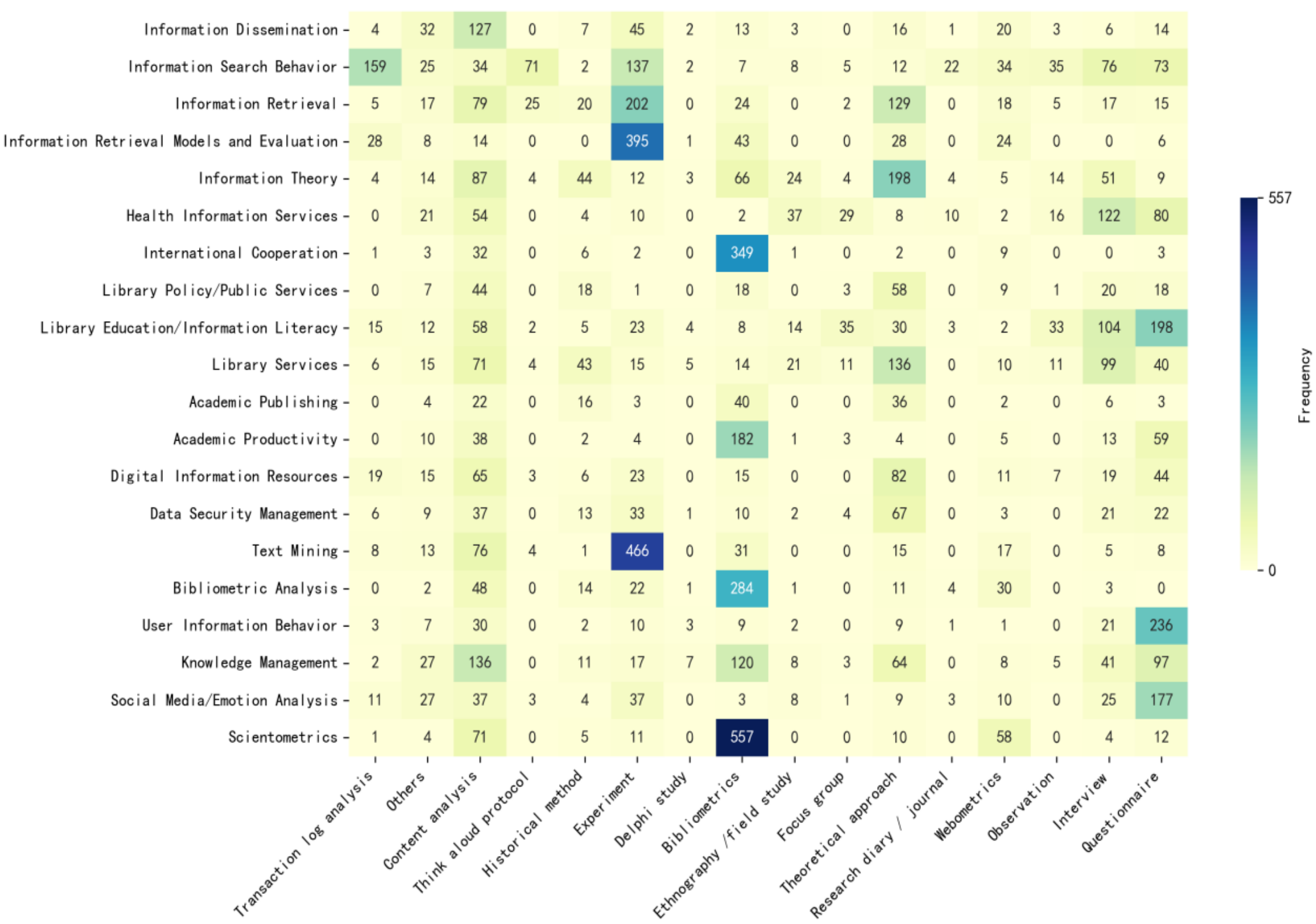

Figure A. 5. The research method of the mid-career scholar uses heat maps.

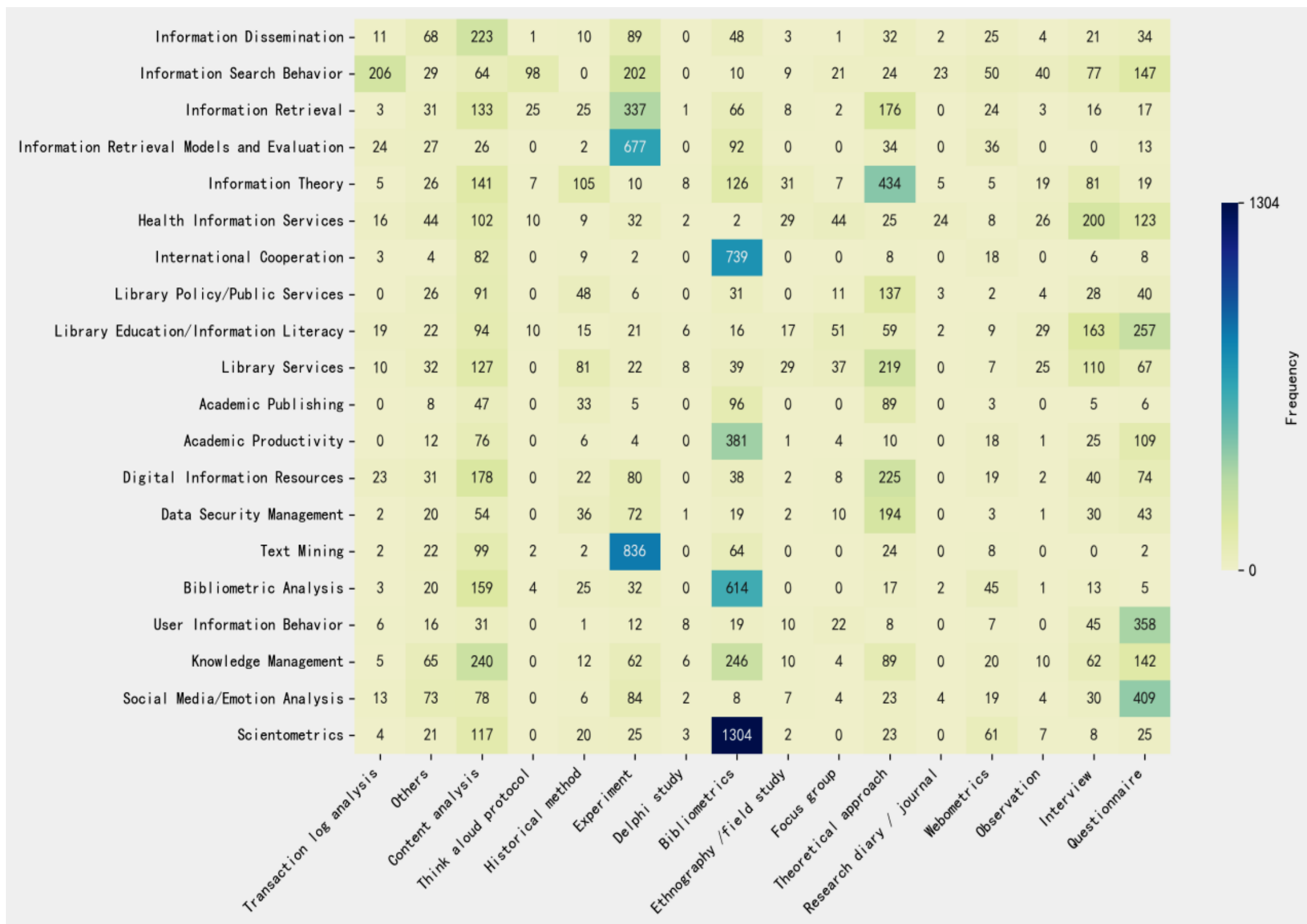

Figure A. 6. The research method of the late-career scholar uses heat maps.

## Appendix. B

Table B. 1. Results of research topic identification in the field of LIS academic papers.

| No. | Topic name | Number | Percentage | No. | Topic name | Number | Percentage |
|---|---|---|---|---|---|---|---|
| 1 | Scientometrics | 2449 | 9.18% | 11 | International Cooperation | 1395 | 5.23% |
| 2 | Information Theory | 1821 | 6.83% | 12 | Bibliometric Analysis | 1311 | 4.91% |
| 3 | Information Retrieval | 1662 | 6.23% | 13 | Library Policy/Public Services | 1074 | 4.03% |
| 4 | Text Mining | 1644 | 6.16% | 14 | Health Information Services | 1039 | 3.89% |
| 5 | Library Services | 1583 | 5.93% | 15 | Academic Productivity | 970 | 3.64% |
| 6 | Digital Information Resources | 1531 | 5.74% | 16 | Social Media/Emotion Analysis | 954 | 3.58% |
| 7 | Information Retrieval Models and Evaluation | 1493 | 5.60% | 17 | Academic Publishing | 946 | 3.55% |
| 8 | Information Search Behavior | 1479 | 5.54% | 18 | Information Dissemination | 828 | 3.10% |
| 9 | Knowledge Management | 1471 | 5.51% | 19 | User Information Behavior | 807 | 3.03% |
| 10 | Library Education/Information Literacy | 1436 | 5.38% | 20 | Data Security Management | 784 | 2.94% |